\documentclass[twocolumn,tighten,times]{aastex62}
\usepackage{graphicx}

\newcommand{\src}{G310.6$-$1.6}

\catcode`\@=11
\newcommand{\gapprox}{\mathrel{\mathpalette\@versim>}}
\newcommand{\lapprox}{\mathrel{\mathpalette\@versim<}}
\newcommand{\propapprox}{\mathrel{\mathpalette\@versim\propto}}
\newcommand{\@versim}[2]
  {\lower3.1truept\vbox{\baselineskip0pt\lineskip0.5truept
\ialign{$\m@th#1\hfil##\hfil$\crcr#2\crcr\sim\crcr}}}
\catcode`\@=12

\defcitealias{berthiere12}{B12}

\defcitealias{renaud10}{R10}

\shorttitle{EXPANSION OF G310.6$-$1.6}

\begin{document}

\title{On the Expansion, Age, and Origin of the Puzzling Shell/Pulsar
  Wind Nebula G310.6$-$1.6}

\email{reynolds@ncsu.edu}

\author{Stephen P. Reynolds}
\affiliation{Department of Physics, North Carolina State University, 
Raleigh, NC 27695-8202, USA}

\author{Kazimierz J. Borkowski}
\affiliation{Department of Physics, North Carolina State University, 
Raleigh, NC 27695-8202, USA}

\begin{abstract}

  We present a 142-ks {\sl Chandra} observation of the enigmatic
  combination supernova remnant \src\ consisting of a bright
  pulsar-wind nebula driven by an energetic pulsar, surrounded by a
  highly circular, very faint shell with a featureless, probably
  synchrotron, spectrum.  Comparison with an observation 6 years
  earlier shows no measurable expansion of the shell, though some
  features in the pulsar-wind nebula have moved.  We find an expansion
  age of at least 2500 yr, implying a current shock velocity less than
  about 1000 km s$^{-1}$.  We place severe upper limits on thermal
  emission from the shell; if the shell locates the blast wave, a
  Sedov interpretation would require the remnant to be very young,
  about 1000 yr, and to have resulted from a dramatically
  sub-energetic supernova, ejecting $\ll 0.02 M_\odot$ with energy $E
  \lapprox 3 \times 10^{47}$ erg.  Even a merger-induced collapse of a
  white dwarf to a neutron star, with a low-energy explosion, is
  unlikely to produce such an event.  Other explanations seem equally
  unlikely.
  
\end{abstract}


\section{Introduction}
\label{intro}

Pulsar-wind nebulae (PWNe) showcase the energy emitted by fast pulsars in
some combination of Poynting flux and relativistic particles.  Young
PWNe, still inside their natal supernova remnants (SNRs), also exhibit the
interaction of that relativistic material with the innermost ejecta
from the supernova, thus connecting a range of important astrophysical
phenomena:  the pulsars themselves, the initially cold relativistic wind
carrying a ``striped'' magnetic field, the relativistic termination
shock thermalizing the relativistic-particle population somehow, the
complex post-shock flow involving tori and disks, the outer boundary
which might mark a ``piston'' driving a shock into the inner ejecta, and
the entire shell SNR that the PWN inhabits.

The combination of a full-fledged PWN with observed pulsar surrounded
by a normal-appearing shell SNR is surprisingly rare.  Discovering and
characterizing a new such system marks important progress in
developing a sample from which common properties can be distinguished
from individual differences.  Such a system may be the remarkable
combination object discovered by INTEGRAL as IGR J14003-6326, and
associated with a recently discovered SNR/PWN as G310.6--1.6
\citep[][hereafter \citetalias{renaud10}]{renaud10}.  This object has
features that make it unique among all SNRs in the Galaxy. Here we
report new {\sl Chandra} observations of \src.

\section{G310.6--1.6}

\begin{figure}
\centerline{\includegraphics[width=3.4truein]{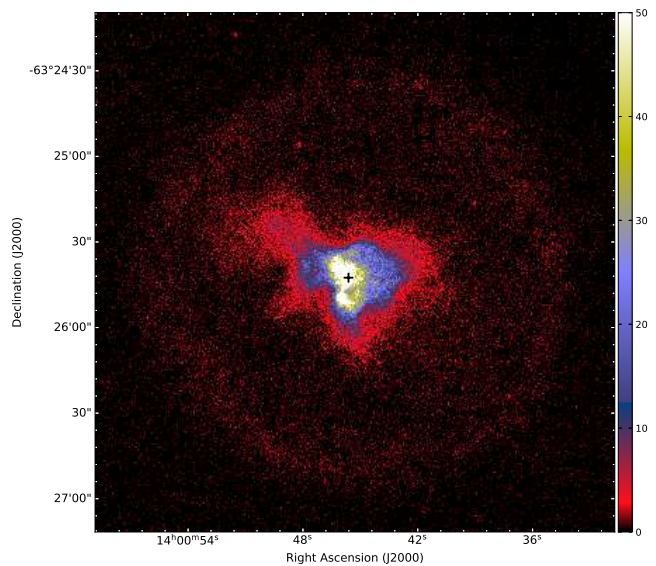}}
\caption{\small{ 142-ks {\sl Chandra} image of G310.6$-$1.6 in 2016 in
    the 1--8 keV energy range (on linear scaling to emphasize its
    shell and the faint structure at the periphery of the PWN). The
    small ($70\arcsec$ in radius) shell is quite spherical but faint,
    with few (up to at most 3) counts per ACIS pixel detected.
    Significant spatial variations within the shell are apparent,
    although there are no obvious very sharp filaments as often found
    in other young SNRs. The central cross indicates the position of
    the pulsar. }}
\label{fig-shell}
\end{figure}

The source IGR 14003-6326 was discovered in a deep INTEGRAL survey of
the Circinus region \citep{keek06} and identified as a small ($\theta
\lapprox 1'$), highly absorbed ($N_H \sim 3 \times 10^{22}$ cm$^{-2}$)
PWN in a 5.1 ks {\sl Chandra} observation \citep{tomsick09}.
\citetalias{renaud10} used {\sl RXTE} observations to discover
pulsations with a period of 31.18 ms and a $\dot P$ implying a
rotational energy-loss rate ${\dot E} = 5.1 \times 10^{37}$ erg
s$^{-1}$, making it one of the most energetic pulsars known.  They
obtained spectra for the pulsar itself and the central PWN, and also
reported a very faint almost circular shell surrounding the PWN, based
on the short {\sl Chandra} exposure.  They christened this previously
unknown shell SNR G310.6--1.6, a designation we shall use for the
entire SNR/PWN system.  See Figure~\ref{fig-shell}.  The PWN had a
hard power-law spectrum, $\Gamma \sim 1.8$, and the shell, while
faint, was well described by a softer power-law with $\Gamma \sim
2.6$.  The distance was uncertain; the dispersion-measure distance of
$10 \pm 3$ kpc would have placed G310.6--1.6 at a $z$-distanpce of 280
pc above the Galactic plane, rather high for a Pop I type object.
However, \citetalias{renaud10} argued that the object might lie in the
Crux-Scutum spiral arm closer to 7 kpc, and used that value in their
analysis.  (We shall adopt the 7 kpc distance, and quote results
in terms of $d_7 \equiv d/7 \ {\rm kpc}$.)
The shell has a very small angular diameter ($\sim 2.5'$),
making it one of the smallest angular-size SNRs in the Galaxy, not
much larger than the youngest Galactic remnant G1.9+0.3.  While
  the PWN is bright, \src\ has not been reported in either GeV or TeV
  gamma rays.

\citetalias{renaud10} interpreted the shell as the outer blast wave of
a normal shell SNR, making this object an interesting new member of
the composite class.  The shell has a diameter of $5.1 d_7$ pc, and if
produced by a $10^{51}$-erg supernova ejecting $10 M_\odot$, has an
initial expansion velocity of 3200 km s$^{-1}$ and an undecelerated
(free expansion) age of only $780 d_7$ yr.  However, from the size of
the PWN relative to the shell and other arguments,
\citetalias{renaud10} concluded that G310.6--1.6 represents the
remnant of a very sub-energetic ($E \sim 5 \times 10^{49}$ erg)
supernova that ejected about 3 $M_\odot$ into a low-density medium
($n_0 \sim 0.01$ cm$^{-3}$).  This would imply an initial expansion
velocity of only 1300 km s$^{-1}$, and a free-expansion age of 1900 yr
(less with deceleration).  Since the spin-down age $P/2{\dot P} =
12.7$ kyr, this interpretation would require the pulsar to have been
born at nearly its current period.

\begin{figure}
\includegraphics[scale=0.31]{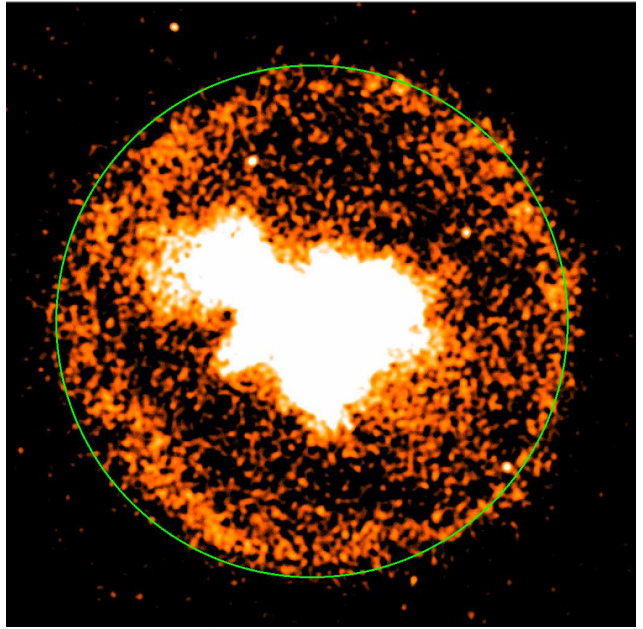}
\caption{\small{\src\ with logarithmic scaling and smoothed with
    a $2\arcsec$ Gaussian.  Note the extension of the PWN toward the shell
    in the NE, possibly reaching as far as the shell, and the clear
    evidence for a dust-scattering halo of emission from the PWN,
    inside the shell.  The circle is centered on the pulsar with a
    radius of $75\arcsec$.}}
\label{smooth}
\end{figure}

However, this interpretation -- in fact, any interpretation -- has
significant problems accounting for all the properties of this
peculiar object.  A 51 ks {\sl Chandra} observation was performed in
2010 (M.~Renaud, PI) but has never been published, though it was
analyzed as part of a bachelor's thesis project at U.~Tours
\citep[][\citetalias{berthiere12}]{berthiere12}.
\citetalias{renaud10} and \citetalias{berthiere12} show that the shell
spectrum is dominantly nonthermal, with no hint of thermal emission.
If this is the blast wave, then, G310.6--1.6 is a member of a very
exclusive club of seven or eight Galactic shell SNRs whose X-ray
spectra are dominated by synchrotron emission.  (See \cite{acero16}
for a current list of SNRs showing synchrotron emission.)  However, in
almost all those instances, and among the slightly larger collection
of SNRs showing some evidence for a synchrotron component in a
dominantly thermal spectrum, the shells show almost unresolved sharp
edges over much of the periphery, certainly not the case for \src.
The large ratio of PWN radius to shell radius here is highly unusual;
there appears to be little room for both shocked surrounding material
and shocked ejecta.  Finally, any interpretation of the shell as a
blast wave requires a very peculiar supernova, as
\citetalias{renaud10} noted.  However, there is no obvious alternative
interpretation of the shell.  It is unlikely to be the PWN/ejecta
interaction shock, as that shock is almost certainly too slow at any
stage of SNR/PWN evolution to be able to accelerate particles to
X-ray-emitting energies (velocities at least of order 800 -- 1000 km
s$^{-1}$ seem required).  Additionally, that would require the
existence of a large completely undetected surrounding shell SNR.

\section{Observations}
\label{obssec}

\src\ was observed in November 2010 (M.~Renaud, PI), and we observed
it in three segments in November and December 2016 (see
Table~\ref{observationlog}).  The two epochs are separated by 6.035
yr. At each epoch, the remnant was placed on the Advanced CCD Imaging
Spectrometer (ACIS) S3 array, and Very Faint mode was used in order to
reduce the particle background. We used CIAO version 4.11 and CALDB
version 4.8.4 to reprocess these observations. No significant particle
flares were found.  To retain the full (subpixel) {\sl Chandra}
  spatial resolution, we used a standard processing option, the
  Energy-Dependent Subpixel Event Repositioning (EDSER) algorithm of
  \citet{li04}, in reprocessed event files.  We aligned all
observations using the bright pulsar, and then merged the three 2016
segments together. The total effective exposure in 2016 is 142 ks,
nearly 3 times as long as in 2010. No statistically significant
displacement was found between the positions of background and
foreground point sources at each epoch (after aligning observations
using the pulsar). (There are only 10 rather faint sources with
matching coordinates, so the alignment based on point sources is
inferior to the alignment based on the pulsar alone.)  We used CIAO
tasks {\tt dmcopy} and {\tt specextract} to extract images and spectra
from the reprocessed and aligned 2010 and 2016 event files at each
epoch.

\begin{deluxetable}{lccc}
\tablecolumns{4}
\tablecaption{{\sl Chandra} Imaging Observations of \src. \label{observationlog}}
\tablehead{
\colhead{Date} & Observation ID & Roll Angle & Effective Exposure \\
& & (deg) & (ks) }
\startdata
2010 Nov 17--18   &  12567 & 151 & 51 \\
2016 Nov 25       &  17905 & 139 & 14 \\
2016 Nov 29--30   &  19919 & 130 & 79 \\
2016 Dec 5        &  19920 & 130 & 49 \\
\enddata
\end{deluxetable}

\section{Imaging}

\subsection{Shell}

Figure ~\ref{smooth} shows a smoothed image, highlighting the shell
and faint emission between the PWN and the shell.  The shell is
remarkably circular; the circle in the figure is centered on the
pulsar, with a radius of $75\arcsec$.  While it is difficult to delineate
the faint outer parts of the PWN, it appears that additional interior
emission is present, expected due to dust scattering from the bright
PWN given the high column density toward \src.  \citetalias{berthiere12}
included such a component in their analysis of radially averaged profiles
of the emission. Beyond the shell, this component forms a low-surface
brightness X-ray halo surrounding the remnant. 

A very rough estimate of the displacement of the pulsar from the
center of a fit of the $75\arcsec$-radius circle to the shell is about
$6\arcsec$, or about $6 \times 10^{17} d_7$ cm.  Even for a very young
age of $700 d_7$ yr, the implied sky-plane speed is only about
$280 d_7$ km s$^{-1}$.

There is a suggestion of emission all the way from the PWN to the
shell in the NE.  If material from the PWN is interacting with the
outer shell, one might expect some spectral signature.  We discuss
this issue below.

\subsection{PWN}

The pulsar in \src\ is bright, with an ACIS-S3 count rate of 0.075 ct
s$^{-1}$ within a circle of radius $4\arcsec$.  The peak rate in the
innermost $3 \times 3$-pixel region is 0.053 ct s$^{-1}$, resulting in
a modest amount of pileup, a reduction of order 8\% of the total count
rate.  This is too low to result in significant image distortion due
to pileup trails.

\begin{figure}
  \centerline{\includegraphics[width=8cm]{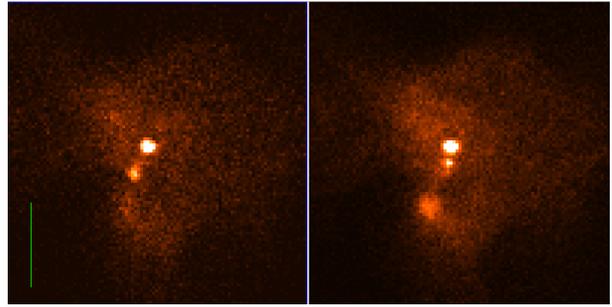}}
    \caption{PWN in 2010 (left) and 2016 (right).  The larger-scale
      structure has changed little, but features just to the south
      of the pulsar have changed markedly.  No clear jet or torus-like
      feature is evident.  The scale bar on the 2010 image has length
    $10\arcsec$.}
    \label{pwn2epochs}
    \end{figure}

\begin{figure}
  \centerline{\includegraphics[width=8cm]{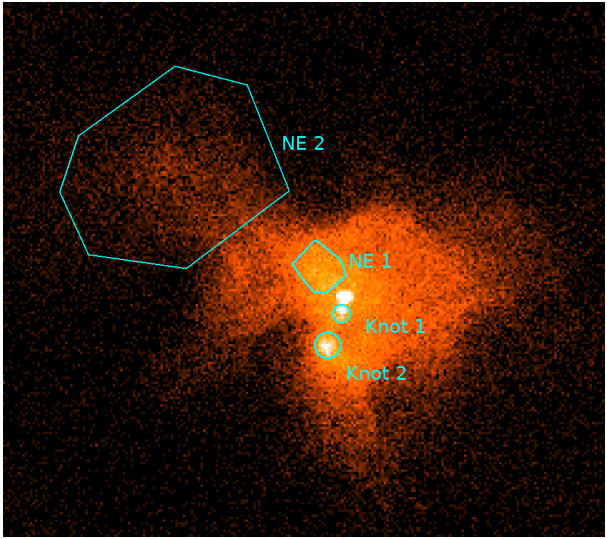}}
  \caption{Regions described in the text, superposed on the 2016
    image.  The pulsar is immediately
  north of Knot 1. Square-root scaling was used.}

    \label{pwnregions}
    \end{figure}

The pulsar-wind nebula has a very similar appearance in both the 2010
and 2016 epochs, with an irregular outline of mean radius about
$30''$, but small structures near the pulsar have changed
(Fig.~\ref{pwn2epochs}).  A bright spot near the pulsar (Knot 1; see
Fig.~\ref{pwnregions}) has changed position by about $1.8\arcsec$
between 2010 and 2016, moving inward toward the pulsar.  If this is a
discrete object, as seems unlikely, the implied speed is about $10,000
d_7$ km s$^{-1}$, but a more likely explanation is a fading of the
first knot and the appearance of a second one.  In the 2016 image, the
knot is separated from the pulsar by about $1.8\arcsec$, so the lower
limit on the required proper motion from the pulsar is the same, about
10,000 km s$^{-1}$, or about 0.03c.  For comparison, knots in the Vela
pulsar-wind nebula are observed to move outward at $0.3 - 0.7 c$
\citep{pavlov03}.  However, spectral fits (see below) indicate that
Knot 1 has not only faded, by about 20\%, but has also changed
spectral slope, increasing the likelihood that two distinct features
are involved at the two epochs.

Knot 2, clearly visible in 2016, was only a faint enhancement of
diffuse emission in 2010.  Unlike Knot 1, it has not moved, but has
brightened by 46\% (see below) while maintaining the same photon
index.

Roughly linear structures can be made out in the PWN at low flux
levels (see Fig.~\ref{pwnregions}), oriented approximately radially.
However, these do not point exactly back to the pulsar position, so they
are not pileup-trail artifacts.

\section{Shell Motion and Age of G310.6$-$1.6}

No shell motion was detected between 2010 and 2016, ruling out a very
young ($< 1000$ yr) age for G310.6$-$1.6. The short ($6.0$ yr) time
baseline in combination with the lack of a sharp outer boundary and
the faintness of the shell prevent us from establishing stringent
constraints on the remnant's age. We briefly describe here our method
used to measure expansion of the shell, and summarize the results
obtained.

We extracted broadband (1--8.5 keV) images at each epoch, $182\arcsec
\times 182\arcsec$ in size, with an image pixel of $0.356\arcsec
\times 0.356\arcsec$, and encompassing the entire remnant. We used
these images to measure the expansion of the shell with the help of a
Bayesian method we employed previously in our study of the expansion
of fast supernova ejecta in the very young remnant G330.2+1.0
\citep{borkowski18}. Our chosen image pixel is about half of 
the ACIS pixel in its surface area. This is still not enough to fully sample 
{\sl Chandra'}s PSF near the optical axis and take full advantage of
the EDSER algorithm, but the mean number of counts per image pixel in the
shell drops even further when images with still smaller pixel
sizes are extracted from the event files. There is no advantage in using
such extremely photon-starved images with our expansion measurement method.

\begin{figure}
   \includegraphics[width=3.5truein]{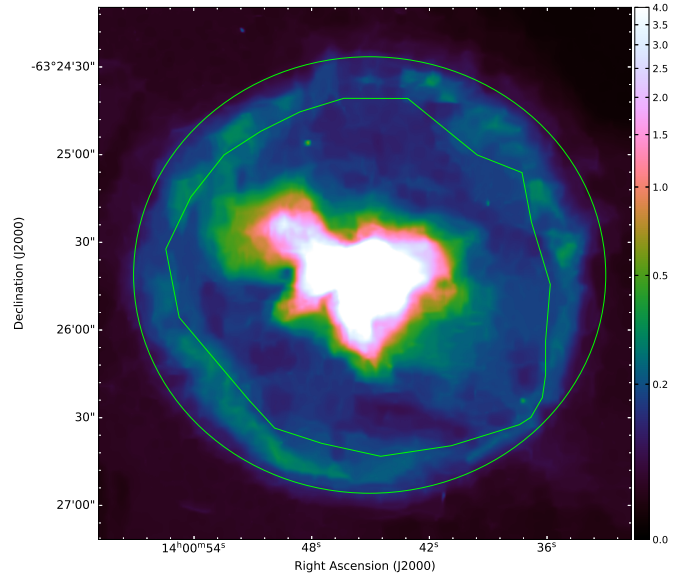}
   \caption{Smoothed {\sl Chandra} image of G310.6$-$1.6 (1--8.5 keV), with
   the region shown where expansion of the shell was measured. The pulsar and
   the inner part of the PWN are saturated. Scale is in counts per
   $0.356\arcsec \times 0.356\arcsec$ image pixel.}
\label{g310smoothedfig}
\end{figure}

The 2016 image was smoothed with the iterative variance-stabilization
method of \citet{azzarifoi16}. The smoothed image is shown in
Figure~\ref{g310smoothedfig}, with the shell region where we measured
expansion overlaid. The 2010 image did not require smoothing.  We
subtracted background from the smoothed 2016 image, and then corrected
the background-subtracted image with a monochromatic ($E = 2.7$ keV)
exposure map prior to expansion measurements. The background was
modeled as a sum of a uniform background and an X-ray halo visible
beyond the remnant's outer boundary. This halo likely arises from
scattering of X-rays produced by the pulsar and the PWN by dust along
the line of sight to G310.6$-$1.6. The background-subtracted and
exposure-corrected image was then allowed to shrink or expand (and
also vary in surface brightness) during Markov chain Monte Carlo
(MCMC) simulations using the PyMC software package
\citep{patil10}. The background in 2010 was also modeled as a sum of a
uniform background and the dust-scattering halo model just
described. Poisson statistics was assumed for the unsmoothed 2010
image.

Expansion might not be centered exactly at the geometrical shell
center even when an SNR shell appears nearly spherically symmetric
\citep[e.g.,][]{williams13}. In order to allow for this possibility,
we allowed the expansion center to vary in our MCMC simulations. A 2-D
Gaussian prior for the expansion center was assumed, centered at the
geometrical shell center, with a FWHM of $16.5\arcsec$ (i.e., $1\sigma
= 7\arcsec$, one tenth of the remnant's radius). With this width, the
allowable displacement of the expansion center exceeds the pulsar's
displacement from the geometrical shell center. Uniform priors were
used for the expansion and the surface brightness scaling factor.

MCMC simulations involved running 10 chains with 6000 samples each,
with each thinned by a factor of 5. Only the results for the expansion
rate are of interest here. (The surface brightness scaling factor does
not differ significantly from unity once {\sl Chandra}'s systematic
flux calibration errors are taken into account, while the expansion
center location cannot be meaningfully constrained in the absence of
discernible shell motion.) The mean expansion rate is negative,
$-0.04\%$ yr$^{-1}$, with the minus sign denoting contraction instead
of expansion. However, the $90\%$ credible interval of $(-0.14,
0.04)\%$ yr$^{-1}$ is consistent with no shell motion firmly
detected. The posterior marginal probability distribution for the
expansion rate is asymmetric, skewed towards negative values away from
its maximum near $-0.025\%$ yr$^{-1}$.

Only the statistical errors arising from the counting noise in the
2010 image were taken into account in our MCMC simulations, so the
errors quoted above underestimate the true errors. The counting noise
in the 2016 image is relatively modest in comparison, leading to an
increase in errors by about $16\%$, but other sources of error (e.g.,
such as arising from smoothing) have not been taken into account.  So
the true $90\%$ credible interval for the expansion rate must be
somewhat wider than obtained from our MCMC simulations. Nevertheless,
fast expansion rates $\ge 0.1\%$ yr$^{-1}$ (corresponding to free
expansion ages $\le 1000$ yr) are strongly disfavored. Instead, our
expansion measurements indicate that G310.6$-$1.6 is most likely at
least several thousand years old.

\section{Spectroscopy}

\begin{figure}
  \hspace*{-5mm}
  \includegraphics[width=6.5truecm,angle=270]{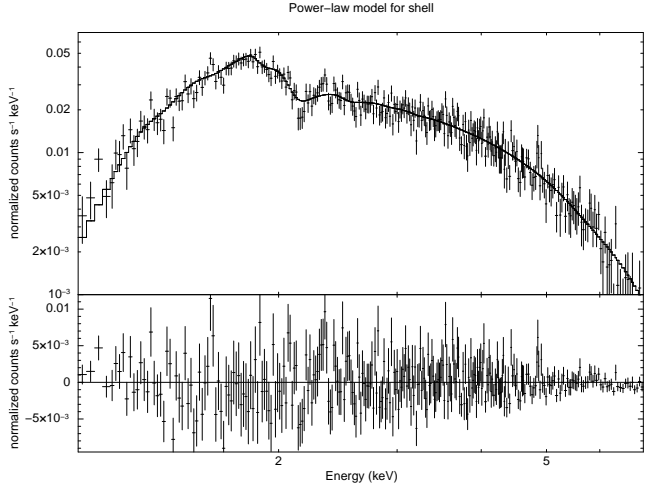}
  \caption{Shell spectrum, fit with power-law model.  See text for details.}
  \label{shellspectrum}
  \end{figure}

We analyzed our combined 142 ks total 2016 observation using XSPEC
version 12.10 \citep{arnaud96}, with \cite{grsa98} abundances and
\cite{cash79} statistics.  (Derived values for column densities can be
significantly different when different abundance sets are used.)
Unless otherwise noted, all fits were performed in the energy range of
0.8 to 7 keV, and all uncertainties are 90\% confidence intervals.
The background was modeled, not subtracted while fitting spectra,
  but only background-subtracted spectra are shown in Figures
  \ref{shellspectrum}, \ref{figshellspectra1}, and
  \ref{figshellspectra3}. \edit3{The particle background was modeled with a
combination of Gaussians and power-laws, exponentially cut off at both
low and high energies. Sky background near \src\ includes the
dust-scattering halo, which we modeled with an absorbed power law.}

\subsection{Spatially-integrated Shell Spectrum}

We confirm with much better statistics the absence of spectral lines
reported by \citetalias{renaud10} and \citetalias{berthiere12},
supporting synchrotron emission as the most likely emission mechanism.
See Figure~\ref{shellspectrum}. The fit shown there reports a column
density $N_{\rm H} = 2.75 (2.63, 2.88) \times 10^{22}$ cm$^{-2}$ and
photon index $\Gamma = 2.55 (2.47, 2.64)$.  The (absorbed) flux
between 0.8 and 7 keV is $1.1 \times 10^{-12}$ erg cm$^{-2}$ s$^{-1}$;
for this column density, the unabsorbed flux is $2.59 \times 10^{-12}$
erg cm$^{-2}$ s$^{-1}$, for an X-ray luminosity of $1.5 \times 10^{34}
d_7^2$ erg s$^{-1}$.

We also fit the data with a simple {\tt srcut} model (synchrotron
radiation from a power-law distribution of electron energies with an
exponential cutoff at some $E_{\rm max}$).  Apparently a radio
counterpart to the X-ray shell has been detected (W. Robbins 2014,
unpublished PhD dissertation)\footnote{Abstract available at
  https://ses.library.usyd.edu.au/handle/2123/12378} but no details
are available.  Therefore, to constrain the fit, we use the upper
limit to 1 GHz radio emission from the shell from
\citetalias{renaud10}: a surface brightness at 1 GHz of $2.3 \times
10^{-21}$ W m$^{-2}$ Hz$^{-1}$ sr$^{-1}$.  We estimate a solid angle
subtended by the shell of an annulus of inner and outer radii of
$50''$ and $70''$ respectively, or 2.1 arcmin$^2$, and obtain an upper
limit for the 1 GHz flux density of 40 mJy.  With this value fixed, we
obtain an energy spectral index of 0.49 (0.47, 0.50) applying to the
power-law portion of the spectrum, and a rolloff frequency of 1.4
(1.1, 1.8)$ \times 10^{17}$ Hz, quite reasonable values, and
consistent with those found with the smaller datasets in
\citetalias{renaud10} and \citetalias{berthiere12}.

To constrain the possible presence of thermal emission, we performed a
two-component fit of a power-law plus plane shock (XSPEC model {\tt
  pshock}).  Lines are suppressed by a very low ionization timescale
$\tau \equiv n_e t$: $3.1 (0, 20) \times 10^8$ cm$^{-3}$ s, where the
90\% confidence interval includes zero.  Thus the model fit is
essentially a bremsstrahlung continuum, with temperature $kT_e = 2.2
(1.4, 3.7)$ keV. The fitted emission measure corresponds to a pure-H
rms density $n_e = 0.050 (0.028, 0.070) d_7^{-1/2}$ cm$^{-3}$
assuming a volume filling factor of 0.25.  It is clear that the shell
is very faint, and the density very low; the preshock density $n_0$ is
lower by a factor of the compression ratio which we take to be 4: $n_0
= 0.013 (0.007, 0.018) d_7^{-1/2}$ cm$^{-3}$. In our fit,
  the thermal component is assigned about 27\% of the total shell
  counts.  However, the two-component model is not statistically
  preferable to our pure power-law model, so we conservatively treat
  the derived density as an upper limit.

\subsection{Spectral Variations within the Shell}

\begin{figure}
  \includegraphics[width=3.5truein]{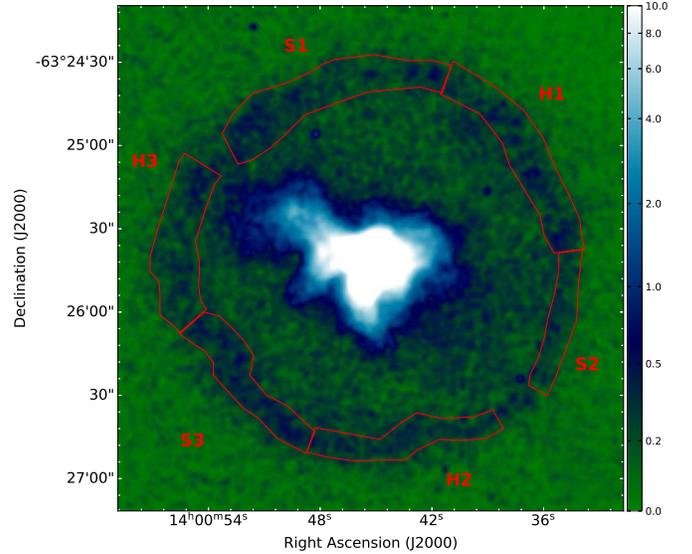}
  \caption{ {\sl Chandra} image of G310.6$-$1.6 (1--8.5 keV; based on
    combined 2010 + 2016 observations), smoothed with
    a Gaussian with the FWHM of $2\arcsec$, and overlaid with shell regions
    chosen for spectral analysis. Scale is in counts per
   $0.356\arcsec \times 0.356\arcsec$ image pixel.}
\label{figshellregions}
\end{figure}

We chose 6 shell regions for detailed spectral studies (see Figure
\ref{figshellregions}). They almost completely cover the entire shell,
with only 2 narrow gaps excluded in order to avoid areas most strongly
affected by out-of-time events originating in the pulsar/PWN
system. The spectral hardness ratio $R = (H-S)/(H+S)$ (where $S$ ($H$)
is the number of detected source counts in the 1--2.5 (2.5--8.5) keV
range) for these regions varies cyclically along the shell: $-7.5\%
\pm 2.5\%$ and $8.8\% \pm 2.7\%$ for regions S1 and H1, $-7.1\% \pm
4.5\%$ ($0.3\% \pm 3.4\%$) for regions S2 (H2), and $-7.9\% \pm 2.9\%$
($4.8\% \pm 3.1\%$) for regions S3 (H3) (errors here are $1 \sigma$).
It is nearly the same among Regions S1, S2, and S3, implying little
(if any) spectral variations between them.  Fits with an absorbed
power law confirm this lack of spectral variations, so we refer to the
set of Regions S1, S2, and S3 as the soft region set S hereafter.
Regions H1, H2, and H3 have significantly harder spectra than the soft
region set S.

\begin{figure}
  \hspace*{-5mm}
  \includegraphics[angle=270,width=3.5truein]{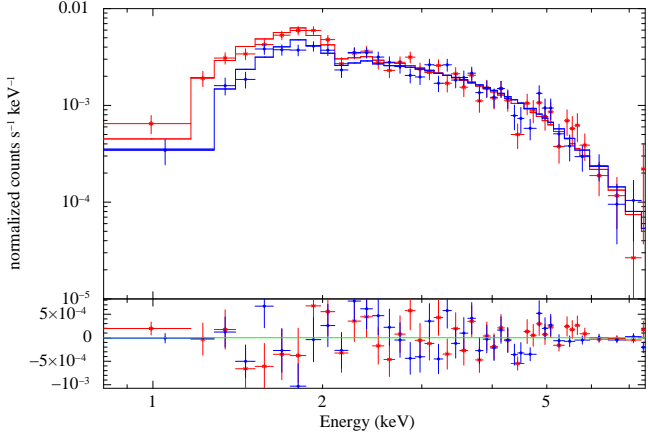}
  \caption{Shell spectra of Regions S1 (red stars) and H1 (blue
    circles). Best-fit models (solid lines) from Table \ref{tableshellfits}
    and residuals (data $-$ model) are also shown. }
\label{figshellspectra1}
\end{figure}

Fits with an absorbed power law give $N_{\rm H} = 2.54 (2.33, 2.76)
\times 10^{22}$ cm$^{-2}$ and $\Gamma = 2.65 (2.50, 2.82)$ for the
region set S, and $N_{\rm H} = 3.65 (3.22, 4.12) \times 10^{22}$
cm$^{-2}$ and $\Gamma = 2.90 (2.63, 3.18)$ for the hardest Region H1.
Since the confidence intervals for $N_{\rm H}$ are well separated, we
conclude that absorption varies significantly across the remnant.  A
joint fit to the region set S and Region H1, assuming the same
intrinsic shell spectrum there but different $N_{\rm H}$, gives
$N_{\rm H} = 2.62 (2.43, 2.82) \times 10^{22}$ cm$^{-2}$ for the
region set S, and $N_{\rm H} = 3.39 (3.13, 3.67) \times 10^{22}$
cm$^{-2}$ for Region H1, with $\Gamma = 2.72 (2.58, 2.86)$. The
hydrogen column $N_{\rm H}$ in Region H1 is $30\%$ larger than within
the region set S. This much larger than average $N_{\rm H}$ in Region
H1 explains why the X-ray spectrum there is harder than elsewhere.
Figure \ref{figshellspectra1} graphically shows spectral differences
between Regions S1 and H1.

\begin{deluxetable*}{cccc}
\tablecolumns{4}
\tablecaption{Spatially-Resolved Shell Models\label{tableshellfits}}
\tablehead{
\colhead{Region} & $N_{\rm H}$ & $\Gamma$ & $K$\tablenotemark{a} \\
& ($10^{22}$ cm$^{-2}$) &  & ($10^{-4}$ ph keV$^{-1}$ cm$^{-2}$ s$^{-1}$) } 

\startdata
S1    & $2.62 (2.43, 2.82)$\tablenotemark{b} & $2.72 (2.58, 2.86)$\tablenotemark{c} & $1.82 (1.51, 2.22)$ \\
S2    & $2.62 (2.43, 2.82)$\tablenotemark{b} & $2.72 (2.58, 2.86)$\tablenotemark{c} & $0.58 (0.48, 0.72)$\\
S3    & $2.62 (2.43, 2.82)$\tablenotemark{b} & $2.72 (2.58, 2.86)$\tablenotemark{c} & $1.34 (1.11, 1.63)$ \\
H1    & $3.39 (3.13, 3.67)$ & $2.72 (2.58, 2.86)$\tablenotemark{c} & $1.96 (1.61, 2.42)$ \\
H2    & $2.43 (2.04, 2.87)$ & $2.37 (2.08, 2.68)$ & $0.69 (0.46, 1.04)$ \\
H3    & $2.98 (2.55, 3.45)$ & $2.56 (2.27, 2.86)$ & $1.21 (0.81, 1.86)$ \\
\enddata
\tablecomments{Errors are $90\%$ confidence intervals.}
\tablenotetext{a}{Normalization of power law at 1 keV.}
\tablenotetext{b}{Assumed the same in Regions S1 -- S3.}
\tablenotetext{c}{Assumed the same in Regions S1 -- S3 and H1.}
\end{deluxetable*}

\begin{figure}
  \hspace*{-5mm}
  \includegraphics[angle=270,width=3.5truein]{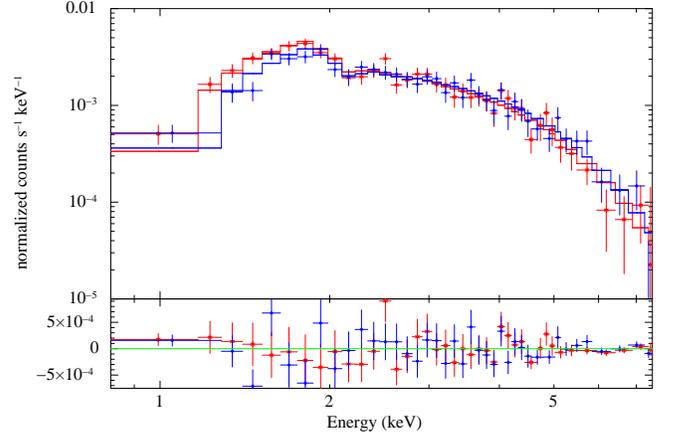}
  \caption{Shell spectra, best-fit models, and residuals for Regions S3
    (in red) and H3 (in blue).
  }
\label{figshellspectra3}
\end{figure}

We also fit spectra extracted from Regions H2 and H3 with an absorbed
power law. Table \ref{tableshellfits} lists results of these fits as
well as the results of the joint spectral fit just described, while
Figure \ref{figshellspectra3} shows how spectra and spectral fits
differ between Regions S3 and H3. In Region H3, the maximum-likelihood
value of $2.98 \times 10^{22}$ cm$^{-2}$ for $N_{\rm H}$ is
intermediate between the region set S and Region H1, while $\Gamma =
2.56$ is slightly less than in these regions. So both high absorption
and an intrinsically hard spectrum might be responsible for the harder
than average spectrum of Region H3.  But the errors on $N_{\rm H}$ and
$\Gamma$ are large enough to make it difficult to distinguish between
high absorption and an intrinsically hard spectrum as the origin for the
harder than average spectrum there. The hardness ratio in Region H3 is
only marginally smaller than in Region H1. A joint fit to their
spectra gives $N_{\rm H} = 3.35 (3.04, 3.68) \times 10^{22}$ cm$^{-2}$
and $\Gamma = 2.77 (2.55, 2.96)$. This best-fit model is remarkably
similar to the best-fit model for Region H1 listed in Table
\ref{tableshellfits}, confirming the spectral similarity of Regions H1
and H3.  Just as for Region H1, high absorption alone can account for
the hardness of Region H3. But in view of large errors for $N_{\rm H}$
and $\Gamma$ (see Table \ref{tableshellfits}), a hard intrinsic
spectrum without any extra absorption remains a plausible alternative
explanation.

Perhaps the most promising hint in favor of intrinsic spectral
variations within the shell comes from Region H2, since the
maximum-likelihood value of 2.4 for $\Gamma$ is the lowest among all
regions (Table \ref{tableshellfits}).  But $N_{\rm H}$ of $2.4 \times
10^{22}$ cm$^{-2}$ is also low, and since errors on $N_{\rm H}$ and
$\Gamma$ are large and they correlate very strongly in our fits, a
higher than average absorption without any intrinsic spectral
variations can also account for the difference in the spectral
hardness of Region H2 and the region set S. By jointly fitting X-ray
spectra from these regions, an increase in $N_{\rm H}$ that is enough
to explain this difference is estimated to be at around $10\%$.

We conclude that substantial ($\sim 30\%$) spatial variations in the
interstellar medium (ISM) column density $N_{\rm H}$ are present along
the line of sight to G310.6$-$1.6, with the largest absorption
detected in the northwest. Patchy absorption with relative variations
of $N_{\rm H}$ up to several tens of percent can explain all spectral
variations detected in its limb-brightened X-ray shell.  Such
structures can be seen in the direction of \src\ in the
DECam Plane Survey \citep{schlafly18}.  But this does not exclude a
possibility that intrinsic spectral variations are as important as
variations in the ISM absorption. In particular, origins of much
harder than average spectra seen in the eastern shell are unclear at
this time. This harder spectrum might be intrinsic to the remnant,
perhaps related to the distorted eastern shell
morphology. Alternatively, the ISM absorption might be nearly as high
there as in the northwest.

\subsection{Pulsar-Wind Nebula}

For the spectrum of the PWN, we also find results consistent with
previous work: $N_{\rm H} = 2.89 (2.85, 2.93) \times 10^{22}$
cm$^{-2}$ and $\Gamma = 2.07 (2.05, 2.10)$.  This value for the photon
index is quite typical for PWNe \citep[e.g.,][]{kargaltsev13}.  The
absorption is marginally higher than that found for the S regions of
the shell, and lower than that in H1 and H3, that is, within the range
of variation we find in the shell.  The unabsorbed flux of the PWN is
$2.20 (2.17, 2.24) \times 10^{-11}$ erg cm$^{-2}$ s$^{-1}$, or $1.3
\times 10^{35}$ erg s$^{-1}$ at a distance of 7 kpc.  This is about
$4 \times 10^{-3}$ of the pulsar spindown power, also a typical fraction.

\citetalias{berthiere12} reported that the spectrum of the PWN softens
with distance from the pulsar, as observed in virtually all PWNe.  At
epoch 2016, the spectrum of Knot 1 near the pulsar at epoch 2016 is
well-described by a featureless power-law with photon index $\Gamma =
1.67 (1.49, 1.85)$, while Knot 2 has $\Gamma = 1.98 (1.84, 2.13)$.  A
more diffuse region in the NE (NE 1 in Fig.~\ref{pwnregions}) has
$\Gamma = 1.87 (1.78, 1.95)$, while fainter emission between the
bright PWN and the shell, Region NE 2, shows a considerably steeper
spectrum, $\Gamma = 2.39 (2.30, 2.49)$, intermediate between the value
in NE 1 and that in the shell.  The difference between NE2 and the
shell is marginally significant; if all regions are required to have
the same column density as that from the fit to the full PWN, we find
$\Gamma({\rm shell}) = 2.64 \pm 0.04$ and $\Gamma({\rm NE2}) = 2.30
(2.26, 2.36)$ -- significantly different.  However, as we find
substantial variations in $N_{\rm H}$ toward different regions of the
shell, which we argue are significant, similar differences toward
different regions of the PWN may also exist.  We can say with
confidence that the photon index of NE2 lies between that of the PWN
as a whole and that of the shell.

In the 2010 observations, both Knots 1 and 2 had substantially
different fluxes (see Table~\ref{knotfluxes}).  Knot 1's photon index
may have changed; the 2010 values are 1.91 (1.67, 2.17), marginally
consistent; for Knot 2, the values are very similar: 1.97 (1.84,
2.10).  It is likely that the feature we label Knot 1 is two features
at the two epochs.  These changes have similar magnitudes to those
observed in some other PWNe, such as Vela \citep{hui17}.

\begin{deluxetable}{lccc}
\tablecolumns{3}
\tablecaption{Fluxes of PWN knots \label{knotfluxes}}
\tablehead{
\colhead{Epoch} & Knot 1 & Knot 2}

\startdata
2010 & 6.66 (5.81, 7.83)  &  5.40 (4.69, 6.39) \\
2016 & 4.76 (4.35, 5.27)  &  8.33 (7.68, 9.09) \\
\enddata
\tablecomments{Fluxes are unabsorbed, in units of $10^{-13}$ erg cm$^{-2}$
  s$^{-1}$ between 0.8 and 7 keV.  Errors are 90\% confidence.}

\end{deluxetable}

\section{Results and Discussion}

Here we enumerate our results:

\begin{enumerate}
  
  \item We find no measurable expansion of the shell in \src\, with a
    lower limit on the expansion age (90\% confidence) of 2500 yr,
    independent of distance.  For a distance of 7 kpc, this
    corresponds to a current shock velocity of no more than 1000 km
    s$^{-1}$.  If deceleration has occurred, the true age will be less
    than the expansion age.
    
\item We confirm the weakness of any thermal emission in the
  shell, with considerably better statistics than in prior
  observations.  A two-component fit including a plane-shock component
  requires a very low density, both because the emission is faint and
  because a very small ionization timescale is demanded by the absence
  of obvious spectral lines.  We conservatively treat the thermal
  component as an upper limit.

  \item Our power-law fits to the integrated shell spectrum are
    consistent with those of previous studies, confirming that
    synchrotron radiation is the dominant emission
    process.

\item We find spectral variations around the shell, most likely due to
  variations in the absorbing column, though changes in the intrinsic
  power-law slope cannot be ruled out.

\item Faint emission from the complex PWN extends in the NE perhaps as
  far as the shell.  The shell structure in the NE shows suggestions of
  interaction with the PWN.

\item We detect dust-scattered halo emission from the PWN both
  interior and exterior to the shell.

\end{enumerate}

We can use these results to constrain the remnant parameters, given
different assumptions about the dynamics.  The plane-shock spectral
component of the shell emission gives a preshock density $n_0 \sim
0.013\, (0.007, 0.018) d_7^{-1/2}$ cm$^{-3}$, and swept-up mass
$M_{\rm sw} \sim 0.031\, (0.017, 0.043) d_7^{5/2} M_\odot$ if that
density were uniform inside the present radius of \src.  An age
estimate of $t = \tau/n_e$ gives $200 d_7^{1/2}$ yr for the best-fit
values of $\tau$ and $n_e$; the 90\% limit for both $\tau$ and $n_e$
gives $2 \times 10^9/0.020$ s or about $2300 d_7^{1/2}$ yr, marginally
consisten with the lower limit we find for the expansion age.

One fundamental question concerning this system is the nature of
the shell itself.  With no measurable expansion velocity, it poses
major problems of interpretation.

\subsection{Blast wave?}

We can attempt to construct a Sedov blast wave given these observed
quantities.  The true age is less than the free-expansion age by 0.4,
or at least 1000 yr; then the observed size gives a relation between
explosion energy and preshock density of $E_{51} \lapprox 0.035\, n_0
d_7^5$.  In a Sedov blast wave, the emission-measure
  weighted ionization timescale $\langle \tau \rangle$ is about 0.2
  times the remnant age times the immediate post-shock electron
  density \citep{borkowski01}, which can be compared with the mean
  ionization timescale in a plane-shock model, $\tau/2$.  Our best-fit
  value of $\tau$ then implies a remnant age of $t \sim 2.5 \tau/n_e
  \sim 500 d_7^{1/2}$ yr.  Our 90\% lower limit on density
  increases this to 900 years; pushing $\tau$ to its 90\% upper limit
  allows us to reach the minimum age of 1000 yr.  Then the swept-up
mass is $0.017(n_0/0.017\ {\rm cm}^{-3}) M_\odot d_7^{5/2}$,
giving an extremely low-energy explosion: $E \lapprox 3 \times 10^{47}
d_7^{9/2}$ erg.  The current shock speed can also be estimated
from the fitted temperature $kT_e = 2.2\, (1.4, 3.7)$ keV, giving
$v_{sh} = 1400\, (1100, 1800)$ km s$^{-1}$, assuming temperature
  equilibration between electrons and ions.  In the absence of
  equilibration, the shock speed could be higher, but again, only the
  90\% lower limit assuming equilibration is consistent with the $\sim
  1000$ km s$^{-1}$ upper 90\% limit we infer.  This picture is
roughly self-consistent; the current kinetic energy of expansion
$M_{\rm sw} v_s^2/2$ is $1.7 \times 10^{47} d_7^{9/2}$ erg,
using the density derived from the plane-shock model emission measure,
about twice the expected 0.3 times the explosion energy for a Sedov
blast wave.

In order for this description to be even approximately
self-consistent, the ejecta mass must be much smaller than the
swept-up mass of $0.017 d_7^{5/2} M_\odot$. Our requirements on
the explosion that could have produced \src\ under this interpretation
are extreme: an explosion energy 3000 times less than that for a
typical core-collapse supernova, occurring in a region of far lower
density than typical in the ISM.  Even for the most extreme possible
distance, $d_7 = 2$, the supernova energy is less than $6 \times
10^{48}$ erg.  A qualitatively similar but less quantitatively extreme
conclusion was reached by \citetalias{renaud10} based on more general
considerations.  For an age of 1000 yr, the pulsar (at its current
luminosity, since the spindown age is much longer) has injected about
$10^{48}$ erg, even more than the explosion energy.  The supernova
explosion energy is normally far larger than the pulsar-injected
energy inferred for young PWNe; the fact that this is not the case
here may be responsible for the unusually large PWN radius compared to
the shell radius.  If \src\ is really in the Sedov phase, this also
implies that the reverse shock has already returned to the center and
recompressed the PWN, an event normally assumed to occur only after
the pulsar luminosity has declined far below its birth value.
However, the required characteristics of the explosion have no
parallel in the study of SNRs.

\subsection{Reverse shock?}

By contrast, it is imaginable that the emission we see is dominantly
from the inward-facing reverse shock, allowing an ejected mass
comparable to rather than much smaller than the swept-up mass.  The
lack of observed expansion would imply that the self-similar
ejecta-driven phase is ending and the reverse shock will be
accelerating back toward the center of the remnant.  (In fact, our
nominal best value for the shell movement is a contraction at 0.04\%
yr$^{-1}$, about 1000 km s$^{-1}$ inward.)  This explanation also
comes at a high price: we must attribute the synchrotron emission to
electrons accelerated at the reverse shock, a phenomenon not clearly
observed in any other remnant.  (However, very few remnants can be
confidently assigned to the relatively brief evolutionary stage in
which the reverse shock is moving back toward the remnant center with
its peak strength.)  The energetics arguments are still constraining:
the observable kinetic energy is orders of magnitude less than the
energy of typical supernovae.  The absence of a clear torus-jet
  morphology as often seen in young PWNe might be taken as an argument
  against a scenario in which the reverse shock has not yet returned
  to crush the PWN, but the diversity of observed PWN morphologies and
  the clear unique features of \src\ suggest that it would be
  premature to rule out any explanations on this basis.

\subsection{Pulsar-fed emission?}

An even more speculative solution attributes the shell emission not to
any shock wave directly, but somehow to particles and magnetic field
injected by the pulsar, perhaps through the PWN which appears to
extend all the way to the inner edge of the shell in the northeast.
There is no precedent for such an interpretation.  High-resolution
radio imaging would be very helpful here, in detecting or
  constraining morphological or spectral connections between the PWN
  and shell.  Consistency, or contrast, of spectral-index or
  polarimetric properties, or particular morphological structures,
  could give clues to the relation.

\subsection{Low-energy supernovae}

All possibilities have in common the requirement of a very low-energy
event (by supernova standards) giving rise to the observed pulsar and
PWN.  One avenue for a weak explosion leaving behind a neutron star is
accretion-induced collapse (AIC) of a white dwarf in a binary.
\cite{dessart06} describe the electron-capture core collapse and
subsequent explosion (ECSN) of a massive white dwarf, ejecting a few
$\times 10^{-3} M_\odot$ with a kinetic energy of $(2-3) \times
10^{49}$ erg, and leaving a neutron star.  The obvious problem with
such an origin for \src\ is the lack of a binary system at present;
unless the mass donor is completely accreted (``merger-induced
collapse''), the supernova will not be adequate to unbind the system.
Furthermore, the particular models of \cite{dessart06} result in
highly asymmetric explosions, at odds with the very circular shell we
observe.  Finally, the rates of such unusual events are expected to be
low \citep{ruiter19}.  However, an anomalously faint supernova, SN
2008ha, has been interpreted as an ECSN, ejecting 0.1 -- 0.3 $M_\odot$
with a kinetic energy of $(1-5) \times 10^{49}$ erg \citep{valenti09}.
Even these extreme examples of weak SNe fail to reach the requirements
for \src.

The PWN would currently be in rough pressure balance with the interior
of our extreme low-energy Sedov blast wave model.
\citetalias{renaud10} report a minimum equipartition energy of about
$4 \times 10^{46} d_7^{17/7}$ erg, implying a pressure $E_{\rm
  min}/3V$ of about $10^{-10} d_7^{-4/7}$ dyn cm$^{-2}$, The
Sedov central pressure $\sim 0.3 \rho v^2$ is about $5 \times 10^{-11}
d_7^{3/2}$ dyn cm$^{-2}$, using the lower limit to the
thermal-gas density in the shell.  Now $E_{\rm min}$ is much less than
the energy injected by the pulsar, so much of that energy has
evidently been radiated.

Our upper limit for the current shock velocity of about $1000 d_7$ km
s$^{-1}$ still allows for ongoing electron acceleration to X-ray
emitting energies, where the rule of thumb requires a minimum shock
speed of about 1000 km s$^{-1}$.  In fact, the rolloff frequency of
about $1.2 \times 10^{17}$ Hz can easily be reached in 1000 years, for
electron acceleration limited either by the remnant age or by
synchrotron losses.  According to expressions in \cite{reynolds08}:
ignoring shock obliquity effects and defining $v_8 \equiv v({\rm
  shock})/10^8\ {\rm cm\ s}^{-1}$, we have

\begin{eqnarray*}
  h\nu_{\rm roll}({\rm age}) & \sim & 0.006\, v_8^4\, \left( t \over 1000\ {\rm yr} \right)^2 \left(B \over 10 \ \mu{\rm G} \right)^3 \ {\rm keV}\\
  h\nu_{\rm roll}( {\rm loss}) & \sim & 0.3\, v_8^2\, \ {\rm keV}.\\
\end{eqnarray*}

The operative value is the lower, so for the observed value of $h
\nu_{\rm roll} \sim 0.5$ keV, a magnetic field of $\sim 56\, \mu$G
would allow acceleration to this value, with acceleration age-limited;
if the actual shock velocity is lower than our upper limit of 1000 km
s$^{-1}$, a higher field would be required, but this is conceivable,
given some amplification as inferred for young SNRs \citep[e.g.,
][]{ressler14}.  But if $B$ is much larger than this, losses would
limit electron energy, requiring $v \sim 1300$ km s$^{-1}$, perhaps
inconsistent with our limit, though more careful calculations would be
required for these estimates to be certain.

\begin{figure}
\vspace{4mm}
  \centerline{\includegraphics[width=8.0cm]{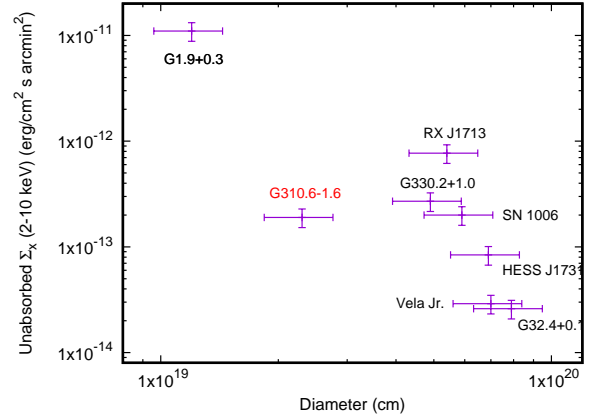}}
  \caption{X-ray surface brightness-diameter plot for 8 Galactic synchrotron
    X-ray-dominated supernova remnants.
    Data: G1.9+0.3, RX J1713--3946, SN 1006, \& Vela Jr.,
    \cite{nakamura12}; G330.2+1.0, \cite{torii06}; HESS J1731--347,
    \cite{doroshenko17}; \& G32.4+0.1, \cite{yamazaki06}.  X-ray
    fluxes are between 2 and 10 keV, and generous uncertainties of
    20\% in both mean X-ray surface brightness $\Sigma_x$ and $D$ were
    assumed.  }
  \label{xssnrs}
  \end{figure}

This presumes the shell does represent the supernova blast wave.  If
the shell represents the reverse shock, the shock speed could be
considerably higher, as undecelerated ejecta enter the roughly
stationary shell and could do so at speeds at or above 1000 km
s$^{-1}$, depending on the remnant age.  But this explanation has
other difficulties as mentioned above.  If the shell is not a shock
wave at all but some other kind of structure, even more problems
arise, as particle acceleration to high energies is still required,
but the true blast wave must be elsewhere and invisible.  While the
shell is faint, its mean surface brightness in X-rays is not too far
below the trend seen in the distribution of X-ray synchrotron shells
from Galactic remnants (see Fig.~\ref{xssnrs}).  Producing synchrotron
emission in a slowly expanding ring that is not a shock wave would be
challenging.  Some mechanism for populating a shell with relativistic
electrons from the PWN might be imagined, but such a hypothesis faces
many difficulties.

The PWN in \src\ is anomalously large as a fraction of the shell
radius (if it is a blast wave), compared to other observed SNR/PWN
systems.  However, the estimates of \cite{vanderswaluw01} contain
enough free parameters that the rough value of 0.4 we estimate for
$R_{\rm PWN}/R_{\rm SNR}$ for \src\ can still be accommodated, in the
Sedov phase when the reverse shock has recompressed the PWN.  If the
shell is a reverse shock, those estimates do not apply, and the large
size of the PWN relative to the shell is easily understandable.  In
this explanation, the reverse shock has yet to return and compress the
PWN.

\section{Conclusions}

We have failed to find any measurable expansion of the highly circular
shell of \src, placing a lower limit on the expansion age of about
2500 y, and an upper limit on the current shock velocity of about
$1000 (d/7\ {\rm kpc})$ km s$^{-1}$.  The shell is very faint; its
spectrum is featureless, almost certainly synchrotron emission.  Upper
limits to thermal emission allow us to deduce that if the shell
locates the blast wave, a consistent description of \src\ as a very
young composite remnant in the Sedov phase is barely possible, with an
age of 1000 yr.  However, a very low-energy explosion is then
required, with energy of less than about $3 \times 10^{47} d_7^{9/2}$
erg, and ejected mass much less than $0.02 d_7^{9/2} M_\odot$.  The
large size of the PWN relative to the shell could be a result of the
very rapid deceleration of the blast wave; in 1000 yr, the pulsar
would deposit about $10^{48}$ erg in the PWN, an unusually high
fraction (more than one!)  of the explosion energy.  While this
interpretation has many problems, alternative explanations face
substantial difficulties as well.

Progress on this fascinating object will require additional {\sl
  Chandra} observations.  A longer time baseline, and an exposure at
least equal to the 142 ks we present here, should confirm expansion or
contraction, or put strong limits on any changes, while improving the
photon statistics for spectral analysis.  The anomalies of
\src\ clearly require some unconventional explanations, which might
advance our understanding of SNRs, PWNe, and their interaction.

\acknowledgments We gratefully acknowledge support by NASA through
                 {\sl Chandra} General Observer Program grant SAO
                 GO6-17062X. The scientific results reported here are
                 based on observations made by the {\sl Chandra} X-ray
                 Observatory.

\software{CIAO (v 4.9) \citep{fruscione06}, ChIPS \citep{germain06},
    XSPEC \citep{arnaud96}, Astropy \citep{astropy13}, matplotlib
    \citep{hunter07}, Numpy \citep{walt11}, PyMC \citep{patil10},
    Scipy\footnote{Jones, E., Oliphant, T., Peterson, P., et al.~2001,
      http://www.scipy.org.}, APLpy \citep{robitaille12} }
  
\vspace{5mm}
\facilities{CXO}

\end{document}